\documentclass[aps,prd,twocolumn,titlepage,superscriptaddress,nofootinbib]{revtex4-2}
\pdfoutput=1

\usepackage{graphicx}      % 插入图片
\usepackage{amssymb}       % 额外数学符号
\usepackage{amsmath}       % 数学公式增强
\usepackage{latexsym}      % 更多符号
\usepackage{cancel}        % 数学中的删除线
\usepackage[normalem]{ulem} % 下划线等
\usepackage{url}           % URL 排版
\usepackage{soul}          % 文字高亮等
\usepackage{verbatim}      % 多行注释
\usepackage{multirow}      % 表格合并单元格
\usepackage{mathrsfs}      % 花体字母
\usepackage{float}         % 浮动体控制
\usepackage[dvipsnames]{xcolor} % 彩色（已包含 color 功能）
\usepackage{mathtools}     % amsmath 扩展
\usepackage{slashed}       % 费曼斜线
\usepackage{physics}       % 物理符号
\usepackage{epstopdf}      % EPS 转 PDF
\usepackage{subfigure}     % 并排子图
\usepackage{bbold}         % 黑体数字
\usepackage{wasysym}       % 更多符号
\usepackage{feynmp}        % 费曼图
\usepackage[colorlinks=true, citecolor=blue, linkcolor=blue, urlcolor=blue]{hyperref}
\usepackage{orcidlink}
\usepackage{etoolbox}

\newcommand{\beq}{\begin{eqnarray}}
\newcommand{\eeq}{\end{eqnarray}}

\makeatletter
\newcommand{\myBig}{\bBigg@{1.75}}
\makeatother

\begin{document}
\title{Phase separation seeded by $\mathbb{Z}_2$ and $U(1)$ topological defects from holography}

\author{Zi-Qiang Zhao\orcidlink{0009-0009-7859-3655}}%\email{zhaoziqiang@stumail.neu.edu.cn}
\affiliation{Liaoning Key Laboratory of Cosmology and Astrophysics, College of Sciences, Northeastern University, Shenyang 110819, China}
\author{Zhang-Yu Nie\orcidlink{0000-0001-7064-247X}}\email{niezy@kust.edu.cn}
\affiliation{Center for Gravitation and Astrophysics, Kunming University of Science and Technology, Kunming 650500, China}
\author{Jing-Fei Zhang\orcidlink{0000-0002-3512-2804}}
\affiliation{Liaoning Key Laboratory of Cosmology and Astrophysics, College of Sciences, Northeastern University, Shenyang 110819, China}
\author{Xin Zhang\orcidlink{0000-0002-6029-1933}}\email{zhangxin@neu.edu.cn}
\affiliation{Liaoning Key Laboratory of Cosmology and Astrophysics, College of Sciences, Northeastern University, Shenyang 110819, China}
\affiliation{MOE Key Laboratory of Data Analytics and Optimization for Smart Industry, Northeastern University, Shenyang 110819, China}
\affiliation{National Frontiers Science Center for Industrial Intelligence and Systems Optimization, Northeastern University, Shenyang 110819, China}

\begin{abstract}
We study the interaction between spontaneous symmetry breaking and phase separation dynamics in holography. Using a double-quench protocol, the system first rapidly crosses the critical point and generates topological defects, while a second quench drives the system into a nonlinear unstable regime with spinodal decomposition. We investigate both $\mathbb{Z}_2$ and $U(1)$ symmetric systems, where different types of topological defects emerge during symmetry breaking. We show that topological defects dynamically determine the nucleation sites of phase separation. As the instability grows, the defect cores expand into macroscopic phase-separated domains. Despite the distinct symmetries and topological properties of these defects, both systems exhibit the same universal dynamical behavior, indicating that topological defects can universally serve as dynamical seeds for subsequent phase separation.
\end{abstract}

\maketitle

\section{Introduction}
Nonequilibrium phase transitions are widely present in condensed matter physics \cite{Zurek:1985qw}, statistical physics \cite{Bray:1994zz}, ultracold atoms \cite{Sadler:2006cok}, cosmology \cite{Kibble:1976sj}, and many other systems \cite{Polkovnikov:2010yn}. When a system rapidly crosses a critical point, critical slowing down prevents different regions from establishing long-range order in time, thereby leading to the spontaneous formation of local domains with different orientations of the order parameter. This process is typically accompanied by the generation of topological defects, such as domain walls in $\mathbb{Z}_2$ systems and vortices in $U(1)$ systems. The statistics of topological defects have been extensively studied by the Kibble-Zurek mechanism \cite{Kibble:1976sj,Zurek:1985qw,Zurek:1996sj,Digal:1998ak,Dodd:1998aan,Carmi:2000zz,delCampo:2013nla,Sonner:2014tca,Li:2021jqk,Zeng:2022hut,delCampo:2022lqd,Xia:2024wfq,Ma:2024srb,Yang:2025bsw,Wang:2025swz} and have been experimentally verified in superfluids \cite{Ko_2019}, superconductors \cite{Monaco_2002}, and ultracold atomic systems \cite{Weiler_2008}. On the other hand, phase separation constitutes another important class of nonequilibrium dynamical processes \cite{10.1063/1.1744102,CAHN1961795,Hohenberg:1977ym,Nicklas_2011}. After entering an unstable region, the system spontaneously develops spatially inhomogeneous structures, which gradually evolve into macroscopically phase-separated regions.
% 非平衡相变广泛存在于凝聚态物理 \cite{Zurek:1985qw}、统计物理 \cite{Bray:1994zz}、超冷原子 \cite{Sadler:2006cok}、宇宙学 \cite{Kibble:1976sj} 以及许多其他系统 \cite{Polkovnikov:2010yn} 中。当系统快速穿越临界点时，临界慢化效应使得不同区域无法及时建立长程序，从而导致自发形成具有不同序参量取向的局域区域。这一过程通常伴随着拓扑缺陷的产生，例如 $\mathbb{Z}_2$ 系统中的畴壁和 $U(1)$ 系统中的涡旋。拓扑缺陷的统计规律已由 Kibble-Zurek 机制 \cite{Kibble:1976sj,Zurek:1985qw,Zurek:1996sj,delCampo:2013nla,Sonner:2014tca,Li:2021jqk,Zeng:2022hut,delCampo:2022lqd,Xia:2024wfq,Ma:2024srb,Yang:2025bsw,Wang:2025swz} 广泛研究，并在超流体 \cite{Ko_2019}、超导体 \cite{Monaco_2002} 和超冷原子系统 \cite{Weiler_2008} 中得到了实验验证。 另一方面，相分离构成了另一类重要的非平衡动力学过程 \cite{10.1063/1.1744102,CAHN1961795,Hohenberg:1977ym,Nicklas_2011}。在进入不稳定区域后，系统会自发形成空间非均匀结构，并逐渐演化为宏观尺度的相分离区域。

Although topological defect formation and phase separation dynamics have both been extensively studied, the intrinsic relationship between them remains to be systematically understood. In particular, whether topological defects in a nonequilibrium system that has undergone symmetry breaking can further affect the subsequent phase separation process and what the ensuing dynamical evolution behavior is remain open questions. Since the core regions of topological defects typically correspond to the vanishing of the order parameter, these regions are likely to be where later instabilities first develop. However, direct investigations of the dynamical mechanism of such defect-induced phase separation are still lacking. Understanding how topological defects affect subsequent nonequilibrium instabilities is important because defects are inherently the most inhomogeneous regions generated during the symmetry breaking process.

% 尽管拓扑缺陷形成与相分离动力学都已被广泛研究，但两者之间的内在联系仍有待系统理解。特别是，在经历对称性破缺的非平衡系统中，拓扑缺陷是否会进一步影响后续的相分离过程以及后学动力学演化行为是什么样的，仍然是一个悬而未决的问题。由于拓扑缺陷的核心区域通常对应于序参量消失的位置，这些区域很可能成为后续不稳定性最先发展的地方。然而，目前对这类缺陷诱导相分离的动力学机制仍缺乏直接研究。理解拓扑缺陷如何影响后续非平衡不稳定性是重要的，因为缺陷本质上是对称性破缺过程中产生的最强非均匀区域。

In recent years, increasing evidence has shown that higher-order nonlinear interactions in holographic superconductor models can naturally produce first-order phase transitions and spatially inhomogeneous phases \cite{Franco:2009yz,Gregory:2009fj,Zhang:2021vwp,Zhao:2022jvs,Zhao:2023ffs,Jin:2026lzp}, thereby providing an ideal platform for studying phase separation dynamics in strongly coupled systems \cite{Janik:2015iry,Janik:2017ykj,Attems:2019yqn,Bellantuono:2019wbn,Li:2020ayr,Chen:2022tfy,Zhao:2023ffs,Chen:2024pyy,Zhao:2026eav,Jin:2026lzp,Zhao:2026jqp}. In this paper, we employ the holographic method \cite{Hartnoll:2008vx,Hartnoll:2008kx,Herzog:2010vz} to study the influence of topological defects on the subsequent phase separation dynamics. To realize topological defect formation and phase separation instability, we adopt a double-quench nonequilibrium protocol. The system first rapidly crosses the critical point, leading to the spontaneous formation of topological defects. Subsequently, a further quench of the nonlinear interaction parameter drives the system into the nonlinear unstable regime where nonequilibrium instability occurs. We investigate two classes of systems with $\mathbb{Z}_2$ symmetry and $U(1)$ symmetry, for which the corresponding topological defects are domain walls and vortices, respectively.
%近年来，越来越多的证据表明，全息超导模型中的高阶非线性相互作用可以自然产生一级相变和空间非均匀相 \cite{Franco:2009yz,Gregory:2009fj,Zhang:2021vwp,Zhao:2022jvs,Zhao:2023ffs,Jin:2026lzp}，从而为研究强耦合系统中的相分离动力学提供了理想平台 \cite{Janik:2015iry,Janik:2017ykj,Attems:2019yqn,Bellantuono:2019wbn,Li:2020ayr,Chen:2022tfy,Zhao:2023ffs,Chen:2024pyy,Zhao:2026eav,Jin:2026lzp,Zhao:2026jqp}。在本文中，我们采用全息方法 \cite{Hartnoll:2008vx,Hartnoll:2008kx,Herzog:2010vz} 研究拓扑缺陷对后续相分离动力学的影响。为了实现拓扑缺陷形成和相分离不稳定性，我们采用了一种双淬火非平衡协议。系统首先快速穿越临界点，导致拓扑缺陷的自发形成。随后，对非线性相互作用参数的进一步淬火将系统驱动到发生非平衡不稳定性的非线性不稳定区域。我们分别研究了具有 $\mathbb{Z}_2$ 对称性和 $U(1)$ 对称性的两类系统，它们对应的拓扑缺陷分别是畴壁和涡旋。

We find that in both types of systems, phase separation does not occur randomly in space but instead develops preferentially in the vicinity of topological defects. As the nonlinear instability grows, the low-condensate regions originally produced by the topological defects broaden and gradually evolve into phase-separated structures of finite size. Although domain walls and vortices have completely different topological and geometric properties, the two types of systems exhibit the same dynamical behavior. This indicates that topological defects can serve as universal dynamical seeds for the subsequent phase separation process.
% 我们发现，在两类系统中，相分离并非在空间中随机发生，而是优先在拓扑缺陷附近发展。随着非线性不稳定性的增长，原本由拓扑缺陷产生的低凝聚区域会不断展宽，并逐渐演化为有限尺寸的相分离结构。尽管畴壁和涡旋具有完全不同的拓扑与几何性质，两类系统却表现出相同的动力学行为。这表明，拓扑缺陷能够作为后续相分离过程的普适动力学种子。

\section{Holographic setup and quench protocol}
\label{sec2}
%既然我们的目标是研究对称性破缺和相分离的的耦合效应，我们首先需要一个能同时实现这两种效应的全息模型。全息超导模型天然的就包含对称性破缺的过程，而相分离效应也可以通过添加高阶非线性项来实现。
Since our main goal is to study the coupling effect between symmetry breaking and phase separation, we first need a holographic model that can realize both phenomena simultaneously. The holographic superconductor model naturally contains the process of symmetry breaking, and phase separation can also be achieved by adding higher-order nonlinear terms ~\cite{Zhao:2022jvs,Zhao:2023ffs,Zhao:2026eav,Jin:2026lzp,Zhao:2026jqp}. To demonstrate the universality of the mechanism, we consider both $\mathbb{Z}_2$ and $U(1)$ symmetric systems. The corresponding holographic models are given by
\begin{align}
\mathcal{L}_{N}=&-\frac{1}{4}h(\Psi_N)F_{\mu\nu}F^{\mu\nu}
-\nabla_{\mu}\Psi_N \nabla^{\mu}\Psi_N
-m^{2}\Psi_N^2
\nonumber\\
&-\lambda\Psi_N^4-\tau\Psi_N^6,
\label{LagN}
\end{align}
with $h(\Psi_N)=e^{\alpha\Psi_N^2}$ and
\begin{align}
\mathcal{L}_{C}=&-\frac{1}{4}F_{\mu\nu}F^{\mu\nu}
-D_{\mu}\Psi_C^{\ast} D^{\mu}\Psi_C
-m^{2}\Psi_C^*\Psi_C
\nonumber\\
&-\lambda(\Psi_C^{\ast}\Psi_C)^{2}
-\tau(\Psi_C^{\ast}\Psi_C)^{3},
\label{LagC}
\end{align}
where $D_{\mu}\Psi_C=\nabla_{\mu}\Psi_C-iA_\mu\Psi_C$ is the standard covariant derivative term of the charged scalar field,  and $F_{\mu\nu}=\nabla_{\mu}A_{\nu}-\nabla_{\nu}A_{\mu}$ is the Maxwell field strength. The $\mathbb{Z}_2$ model supports domain walls, while the $U(1)$ model supports vortices. In both systems, topological defects naturally emerge during quenching across the critical point. 
Since we want to study whether topological defects affect the subsequent phase separation process, we need to place the system in a framework that allows for time evolution, and it is better to use the ingoing Eddington metric
\begin{align}
ds^{2}=\frac{L^2}{z^2}\left(-f(z)dt^{2}-2dtdz+dx^{2}+dy^{2}\right),
\end{align}
with $f(z)=1-(z/z_h)^3$. We adopt the following ansatz
\begin{align}
\Psi_N = \Psi_C &=z\psi(t,z,\vec{x})/L,\\
A_\mu dx^\mu&=
A_t(t,z,\vec{x})dt
+A_{\vec{x}}(t,z,\vec{x})d\vec{x}.
\end{align}
Near the AdS boundary, the asymptotic expansions of the fields are
\begin{align}
A_{t}=\mu-z\rho,
\qquad
\psi=\psi^{(1)}+z\psi^{(2)}+\dots,
\end{align}
where $\mu$ and $\rho$ correspond to the chemical potential and charge density of the boundary theory, respectively. Throughout this work, we impose the source free boundary condition $\psi^{(1)}=0$, such that the condensate is given by $\langle\mathcal{O}\rangle=\psi^{(2)}$. In the following calculations, we fix $m^2=-2$, $\alpha=5$, and $L=z_h=1$. Due to the scaling symmetry, the temperature of the system is $T = 3/(4\pi\rho^{1/2})$. In this manuscript, we fix the charge density $\rho$, which corresponds to the canonical ensemble. The full dynamical equations of motion are presented in Appendix~\ref{app:A}. In these two models, the parameters $\lambda$ and $\tau$ are mainly used to control the stability of the system, and a detailed study has been given in Ref.~\cite{Zhao:2022jvs}.

%在这两个模型中，lambda和tau参数主要用来控制系统的稳定性，文章[]中已经给出了详细的研究。

\section{Defect-seeded phase separation in the $\mathbb{Z}_2$ system}
\label{sec3}
\begin{figure}
    \centering 
    \includegraphics[width=\columnwidth]{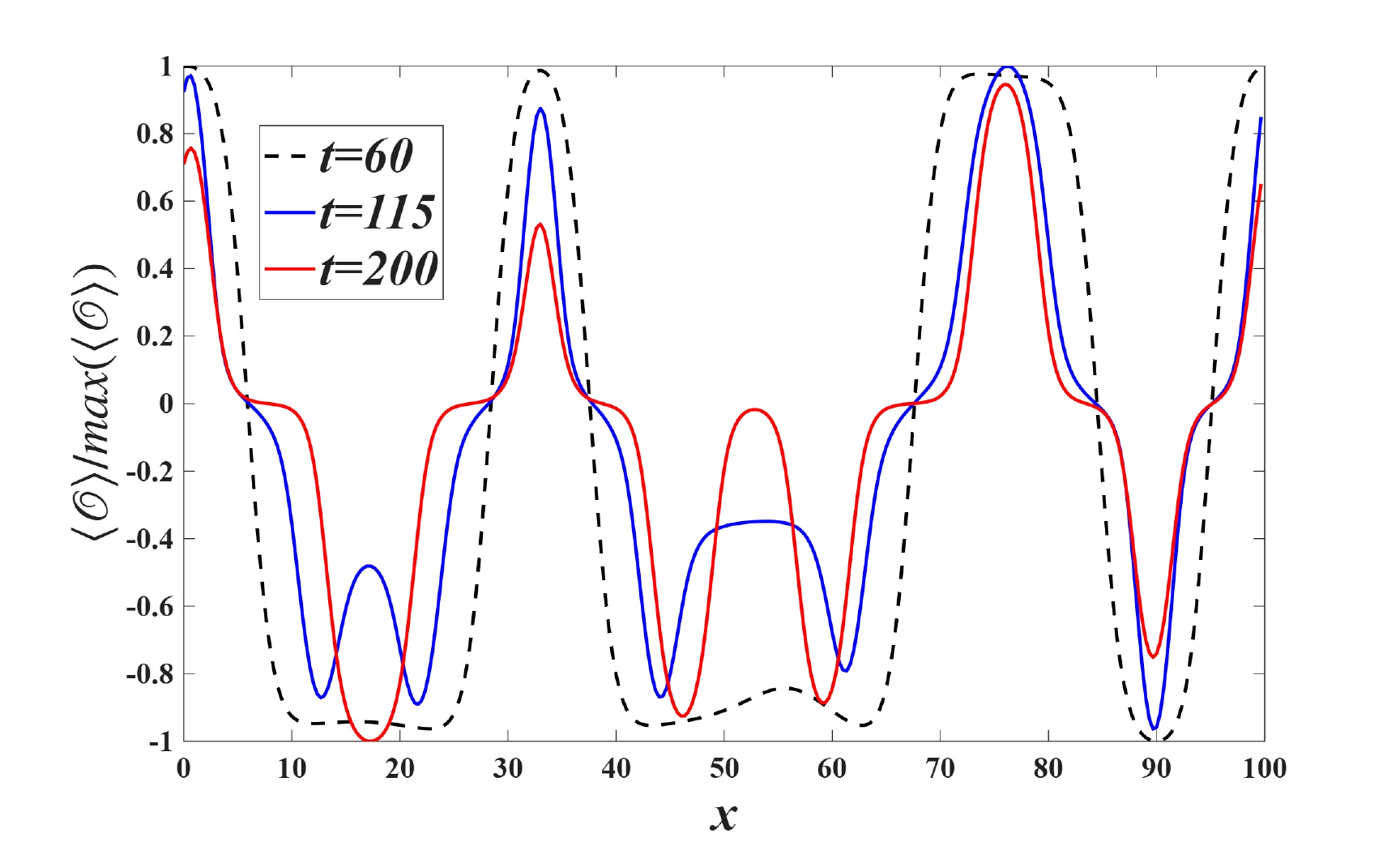}
    \caption{
    The time evolution of the scalar condensate in the one-dimensional double-quench process. 
    The dashed line represents the kink configuration generated during the first quench without phase separation, while the solid line shows the subsequent evolution after $t>100$ when the nonlinear instability is turned on.}
    \label{PhaseSepinvasion1D}
\end{figure}

We start from a simple one-dimensional model. If we set $A_y = 0$, the model reduces to the simplest one-dimensional case, where the topological defects are kinks. The coupling effect between symmetry breaking and phase separation in the one-dimensional case has already been studied in Ref.~\cite{Zhao:2026eav}, but unlike that work, we adopt a new quench strategy. In Fig.~\ref{PhaseSepinvasion1D}, we present such results for the one-dimensional case. 
The quench parameters are $T_i = 1.041 T_c$, $T_f = 0.924 T_c$, and $\tau_Q = 1$. In the $\mathbb{Z}_2$ system, the critical  is $T_c=0.201$. For $t = 0$ to $100$, we have $\lambda = 0$ and $\tau = 3$. In this stage, the system only undergoes symmetry breaking, so the only spatially inhomogeneous structures are kinks. For $t > 100$, we perform a second quench of $\lambda$ at the same rate, such that $\lambda = -4$. At this moment, the system becomes nonlinearly unstable and spinodal decomposition begins.

%对于t>100,我们以相同的速度对lambda进行第二次quench，使得lambda=-4。在这一时刻，系统将存在非线性不稳定性，并且开始进行spinodal分解

\begin{figure*}
    \centering
    \includegraphics[width=2\columnwidth]{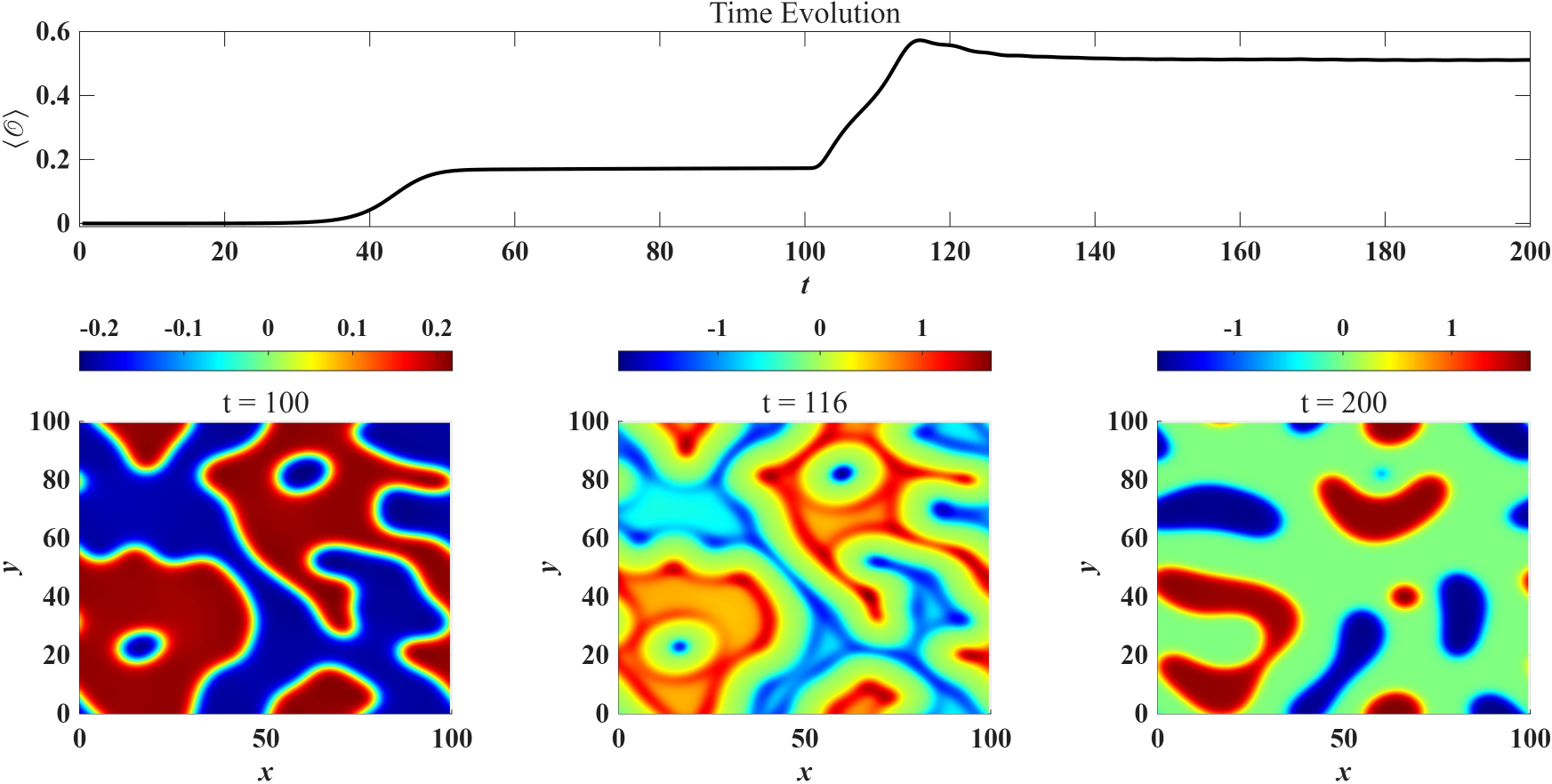}
    \caption{
    Nonequilibrium evolution of the system with $\mathbb{Z}_2$ symmetry.
    The upper panel shows the spatially averaged condensate as a function of time.
    The lower panels display the spatial distribution of the condensate at different times. The color bar indicates the magnitude of the condensate.
    }
    \label{PhaseS2D}
\end{figure*}

%color bar表示凝聚值的大小

Since the spatial inhomogeneity is largest at the kink interfaces, phase separation preferentially occurs at these locations. This result can be clearly observed in Fig.~\ref{PhaseSepinvasion1D}. As the phase separation proceeds, the region where $\langle\mathcal{O}\rangle = 0$ gradually broadens and evolves into an inhomogeneous structure of finite width. The reason for this phenomenon is that bubble expansion requires absorbing energy from the surroundings. The original region where $\langle\mathcal{O}\rangle = 0$ is already in a lower energy state. The phase separation process makes it easier for bubbles to absorb the condensate from these low-energy regions, thereby further expanding the region where $\langle\mathcal{O}\rangle = 0$.

%z2系统二维的情况其实和一维的没有本质上的区别，唯一的不同是二维平面在quench之后产生的拓扑缺陷是domain wall网络，这使得动力学过程看上去会更加复杂。对于二维的情况，我们采取的quench策略和一维的一样，依然是在t<100时进行没有相分离的第一次quench，在t>100时第二次quenc使得tau=-4。正如图PhaseS2D所示，在第一次系统产生了拓扑缺陷，随后第二次quench的时候相分离首先在缺陷的界面处开始发生。同样的，泡泡在膨胀的的过程中会吸收凝聚值比较小的区域的能量，最后导致$\langle\mathcal{O}\rangle = 0$的区域膨胀，这对应于图PhaseS2D中t=200时刻时候的结果。不仅如此，在文章[]中，作者考虑了固定初始扰动情况下的一维的相分离过程，实现了均匀的入侵现象，出于结果完整性考虑，我们同样计算了这种情形下的结果，并且展示在了附录app:B中

The two-dimensional case of the $\mathbb{Z}_2$ system is essentially not different from the one-dimensional case. The only difference is that the topological defects generated after the quench on the two-dimensional plane form a domain wall network, which makes the dynamics appear more complicated. For the two-dimensional case, we adopt the same quench strategy as in one dimension: the first quench without phase separation is performed for $t < 100$, and at $t > 100$ a second quench sets $\lambda = -4$. As shown in Fig.~\ref{PhaseS2D}, topological defects are generated during the first quench. When the second quench is applied, phase separation first occurs at the defect interfaces. Similarly, during bubble expansion, energy is absorbed from regions with a smaller condensate, eventually leading to the expansion of the region where $\langle\mathcal{O}\rangle = 0$, which corresponds to the result at $t = 200$ in Fig.~\ref{PhaseS2D}. Moreover, in Ref.~\cite{Zhao:2026eav}, the authors considered the one-dimensional phase separation process with a fixed initial perturbation and realized a uniform invasion phenomenon. For completeness, we also compute the results in this scenario and present them in Appendix~\ref{app:B}.

%在经历了对称性破缺和相分离之后，系统剩下的过程就是泡泡的合并以及不同拓扑荷之间的相互湮灭。这一过程同样包含了两种不同的机制，这使得其中的动力学更加复杂和难以预测。但是一个比较自然的结果是，小的泡泡会相互合并成大的泡泡，这一点和普通的泡泡动力学是一致的。
After the processes of symmetry breaking and phase separation, the remaining evolution of the system consists of bubble coalescence and mutual annihilation of defects with different topological charges. This process also involves two distinct mechanisms, making the dynamics more complex and difficult to predict. Nevertheless, a natural outcome is that small bubbles merge into larger ones, which is consistent with ordinary bubble dynamics.

\section{Universal defect-seeded phase separation in the $U(1)$ system}
\label{sec4}
An important question is whether the defect-induced phase separation dynamics identified in the $\mathbb{Z}_2$ system remains valid for continuous symmetries. 
In particular, for a system with $U(1)$ symmetry, the topological defects generated during symmetry breaking are vortices rather than domain walls. 
Since vortices are point-like defects carrying nontrivial phase winding, their coupling to phase separation dynamics can, in principle, differ substantially from the $\mathbb{Z}_2$ case.

\begin{figure*}
    \centering
    \includegraphics[width=2\columnwidth]{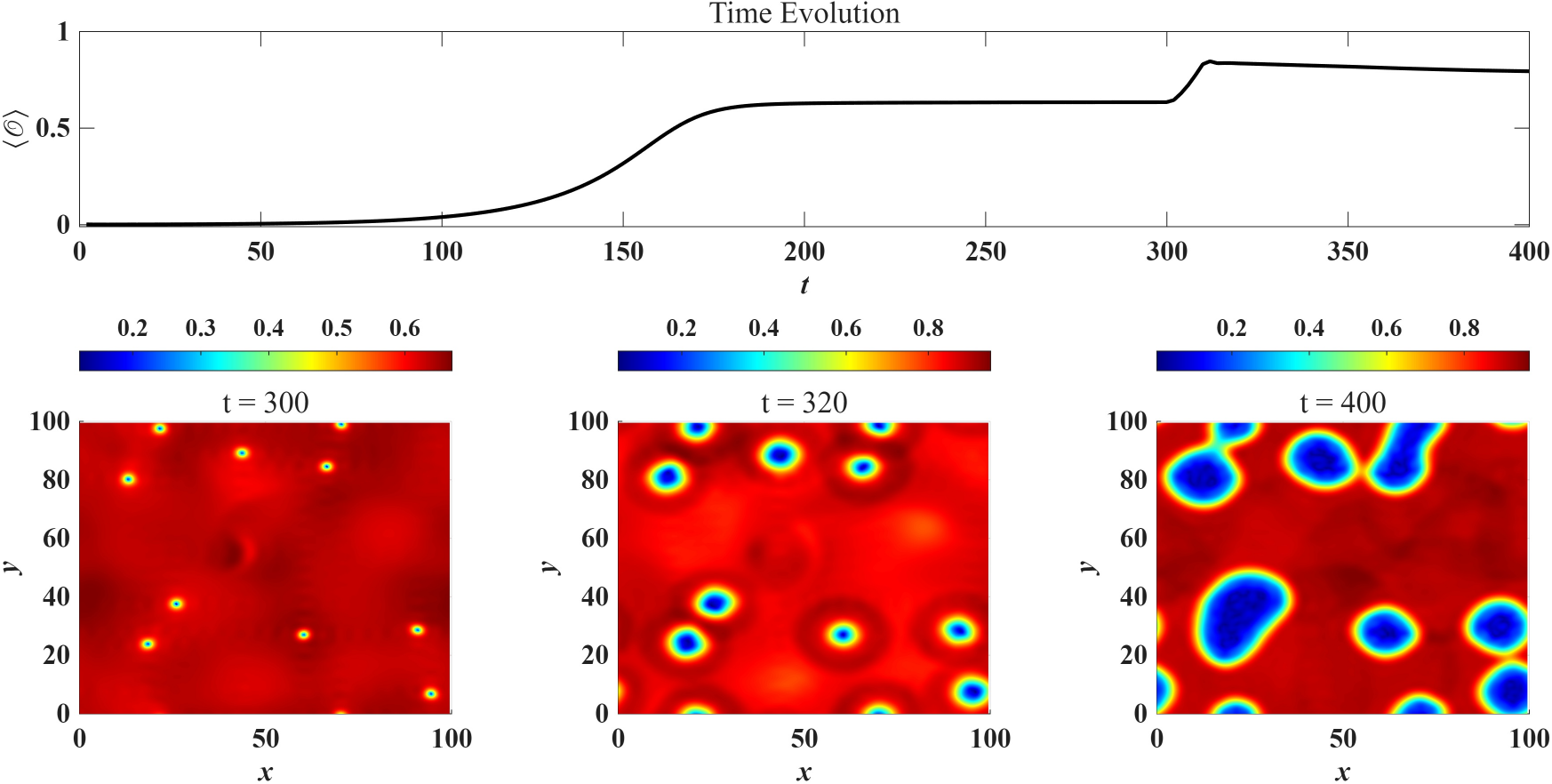}
    \caption{
    Nonequilibrium evolution of the system with $U(1)$ symmetry. The upper panel shows the average condensate as a function of time.
    The lower panels display the spatial distribution of the condensate at different times. The color bar indicates the magnitude of the condensate.
    }
    \label{PhaseS2DVotex}
\end{figure*}

Therefore, we also compute the effect for the system with $U(1)$ symmetry. The dynamical process here is the same as in the $\mathbb{Z}_2$ case, and the double-quench method is also adopted. We first quench the system from the normal phase to the superfluid phase for $t < 300$ with parameters $T_i = 1.164T_c$, $T_f = 0.902T_c$, $\lambda=0$, and $\tau=0.266$, while keeping the quench rate $\tau_Q = 1$ unchanged. In the $U(1)$ system, the critical  is $T_c = 0.118$. During the symmetry breaking stage, multiple vortices and antivortices are generated, accompanied by vortex-antivortex annihilation processes. After the vortex configuration stabilizes, we further quench the nonlinear interaction parameter from $\lambda = 0$ to $\lambda = -1$, which drives the system into the nonlinear unstable regime where spinodal decomposition occurs.
%因此，我们也计算了具有 $U(1)$ 对称性情况下的效应。这里的动力学过程与 $\mathbb{Z}_2$ 情形相同，同样采用了双淬火方法。我们首先在t<300时将系统从正常相快速淬火到超流相，参数为 $T_i = 1.164T_c$ 和 $T_f = 0.902T_c$，同时保持淬火速率 $\tau_Q=1$ 不变。在 $U(1)$ 系统中，临界温度为 $T_c=0.118$。在对称性破缺阶段，会产生多个涡旋和反涡旋，并伴随涡旋-反涡旋的湮灭过程。当涡旋构型稳定之后，我们进一步将非线性相互作用参数从 $\tau= 0$淬火到\tau= -1$ ，这会使系统进入发生旋节分解的非线性不稳定区域。

The complete evolution process is shown in Fig.~\ref{PhaseS2DVotex}. At early times, the system is dominated by vortex-antivortex annihilation. After the second quench, phase separation first occurs at the vortex cores. Since the condensate vanishes at the vortex centers, the originally point-like regions broaden during the evolution and gradually expand into finite-sized phase-separated structures. This process is analogous to the expansion of the $\langle\mathcal{O}\rangle = 0$ region in the $\mathbb{Z}_2$ case. The reason is that bubble expansion requires absorbing the surrounding condensate, and the vortex cores already have zero condensate, making the surrounding regions more easily absorbed by the expanding bubbles. Similar to the $\mathbb{Z}_2$ symmetry case, after bubbles are induced by topological defects, these bubbles expand and collide. Although the geometric structure of topological defects is fundamentally different from the domain walls in the $\mathbb{Z}_2$ system, both types of systems exhibit the same universal dynamical mechanism: phase separation continuously expands the zero-condensate regions created by topological defects.

%完整演化过程如图~\ref{PhaseS2DVotex} 所示。在早期，系统以涡旋-反涡旋湮灭过程为主导。第二次淬火之后，相分离过程首先开始在涡旋处发生。由于凝聚在涡旋中心为零，原本点状区域在演化过程中不断展宽，并逐渐扩展为有限尺寸的相分离结构。这一过程和$\mathbb{Z}_2$中$\langle\mathcal{O}\rangle = 0$区域膨胀的原理一样，因为泡泡膨胀需要吸收周围的凝聚值，而涡旋中心本身凝聚值为零，所以周围的区域更容易被膨胀的泡泡吸收。与 $\mathbb{Z}_2$ 对称性的情形类似，拓扑缺陷诱导出气泡之后，这些气泡会膨胀并发生碰撞。尽管拓扑缺陷的几何结构与 $\mathbb{Z}_2$ 系统中的畴壁有本质不同，但两类系统呈现出相同的普适动力学机制：相分离持续扩展由拓扑缺陷产生的零凝聚区域。

\section{Discussion}\label{sec5}
In this paper, we investigate the influence of topological defects on the subsequent phase separation dynamics and find that topological defects can serve as dynamical seeds for the phase separation process. To simultaneously realize the formation of topological defects and the emergence of nonlinear instability in the strongly coupled regime, we adopt a holographic model with higher-order nonlinear terms and employ a double-quench nonequilibrium protocol during the nonequilibrium dynamical process. The system first rapidly crosses the critical , dynamically generating topological defects via spontaneous symmetry breaking. Subsequently, a further quench of the nonlinear interaction parameter drives the system into the unstable regime where spinodal decomposition occurs.
%在本文中，我们研究了拓扑缺陷对后续相分离动力学的影响，发现拓扑缺陷能够作为相分离过程的动力学种子。为了同时实现强耦合情况下的拓扑缺陷的形成和非线性不稳定性的出现，我们采用了具有高阶非线性项的全息模型，并且在非平衡动态过程中采用了一种双淬火非平衡协议。系统首先快速穿越临界温度，通过自发对称性破缺动态产生拓扑缺陷。随后，对非线性相互作用参数的进一步淬火将系统驱动至发生旋节分解的不稳定区域。

We studied two classes of systems with $\mathbb{Z}_2$ and $U(1)$ symmetry, respectively. In the $\mathbb{Z}_2$ system, the quench generates kink (in one dimension) or domain wall (in two dimensions) structures, while in the $U(1)$ system, the quench leads to vortex-antivortex configurations. Despite their distinct geometric and topological properties, once the system becomes unstable, they all induce the same subsequent dynamics: phase separation develops preferentially around the topological defects, with the initially zero-condensate regions broadening over time. Hence, when phase separation follows symmetry breaking, it is not random but organized by pre-existing defect structures.
%我们分别研究了具有 $\mathbb{Z}_2$ 和 $U(1)$ 对称性的两类系统。在 $\mathbb{Z}_2$ 系统中，淬火产生扭结(一维)或畴壁(二维)结构，而在 $U(1)$ 系统中，淬火之后系统形成涡旋-反涡旋构型。尽管这些拓扑缺陷在几何和拓扑性质上存在显著差异，但是我们发现它们在系统存在不稳定性之后，都会诱导出相同的后续动力学行为：相分离优先在拓扑缺陷周围发展，且缺陷原本产生的零凝聚区域在演化过程中不断展宽。因此，如果系统同时包含了对称性破缺和相分离，相分离并非在系统中随机发生，而是由预先存在的拓扑缺陷结构所组织和引导。这一机制同时出现在不同对称性和不同拓扑结构的系统中，表明缺陷诱导的相分离动力学在一定程度上具有普适性。

More broadly, this work establishes a dynamical connection between spontaneous symmetry breaking and phase separation in strongly coupled systems. In the two-step quench process, the topological defects formed first provide natural seeds for the subsequent occurrence of phase separation. This observation indicates a dynamical link between the topological defects generated by spontaneous symmetry breaking and the phase separation process. In more general cases where symmetry breaking and phase separation occur almost simultaneously, on the one hand the spontaneously generated topological defects become preferred nucleation sites for phase separation. On the other hand, small bubbles produced by phase separation may contract and form topological defects. Therefore, the dynamical interplay between spontaneous symmetry breaking and phase separation may have important consequences for statistical outcomes in nonequilibrium processes, such as the Kibble-Zurek mechanism.
%更广泛地说，这项工作在强耦合系统的自发对称性破缺与相分离之间建立了一种动力学联系。在两步quench过程中，首先形成的拓扑缺陷会为后续相分离的发生提供天然种子。这一现象表明自发对称性破缺产生的拓扑缺陷与相分离过程之间存在动力学关联。在更一般情况下，如果对称性破缺过程和相分离过程几乎同时发生，那么一方面自发产生的拓扑缺陷会成为相分离更易发生的触发点；另一方面，相分离产生的小泡泡也可能在收缩过程中形成拓扑缺陷。因此对称性自发破缺过程和相分离过程之间的动力学关联可能对非平衡过程中的统计结果产生重要影响，比如Kibble-Zurek机制等。

Moreover, since topological defects can dynamically determine the nucleation sites of phase-separated structures, the defect-seeded mechanism revealed in this paper may also provide a new route for actively manipulating spatial composition distributions in superfluid systems via defect engineering. From this perspective, such effects may be experimentally realized and verified in ultracold atoms, multicomponent condensates, and other strongly nonequilibrium quantum fluids in the future.
%另一方面，由于拓扑缺陷能够动态决定相分离结构的成核位置，本文所揭示的缺陷诱导机制也可能通过缺陷为主动操控超流系统中的空间组分分布提供一条新途径。从这个角度看，此类效应有望在未来于超冷原子、多组分凝聚体以及其他强非平衡量子流体中实现实验验证。

\section*{Acknowledgement}
ZQZ thank Xin Zhao for useful discussions. This work is supported by the National Natural Science Foundation of China (grant nos. 12575054, 12533001, 12575049, 12473001, 12205039, 12305058, and 11965013). ZYN is partially supported by Yunnan High-level Talent Training Support Plan Young $\&$ Elite Talents Project (grant no. YNWR-QNBJ-2018-181). This work is also supported by the National SKA Program of China (grant nos. 2022SKA0110200 and 2022SKA0110203) and the 111 Project (grant no. B16009).

\bibliographystyle{apsrev4-1}
\bibliography{reference}

% Appendix in single-column format
\newpage
% 附录部分
\onecolumngrid  % REVTeX 专用的单栏切换命令

% \appendix
% \section{Appendix A: Full Equations of Motion}
% \label{app:equations}

\appendix
\section{Full equations of motion}
\label{app:A}
% 重新设置公式编号为 A-1, A-2, ...
\renewcommand{\theequation}{A-\arabic{equation}}
\setcounter{equation}{0}
% 重新设置图形编号为 Figure A-1, Figure A-2, ...
\renewcommand{\thefigure}{A-\arabic{figure}}
\setcounter{figure}{0}

The formulas for the nonequilibrium evolution of a system with $\mathbb{Z}_2$ symmetry are
\begin{align}
\frac{\partial _z\psi  f'}{2}+\frac{\psi  f'}{2 z}-\frac{1}{4} \alpha  \psi  z^2 e^{\alpha  \psi ^2 z^2} \big(f
   \partial _zA_x{}^2+f \partial _zA_y{}^2-2 \partial _tA_x \partial _zA_x-2 \partial _tA_y \partial _zA_y+2 \partial
   _xA_t \partial _zA_x+\partial _xA_y{}^2-2 \partial _xA_y \partial _yA_x\nonumber\\
   +2 \partial _yA_t \partial _zA_y+\partial
   _yA_x{}^2-\partial _zA_t{}^2\big)
   +\frac{f \partial _z\partial _z\psi }{2}-\frac{f \psi }{z^2}-\lambda  \psi
   ^3-\partial _t\partial _z\psi +\frac{\partial _x\partial _x\psi }{2}+\frac{\partial _y\partial _y\psi
   }{2}-\frac{3}{2} \tau  \psi ^5 z^2+\frac{\psi }{z^2}=0,&\label{A1}\\
2 \alpha  \partial _x\psi  \partial _zA_x \psi  z^2+2 \alpha  \partial _y\psi  \partial _zA_y \psi  z^2-2 \alpha 
   \partial _zA_t \partial _z\psi  \psi  z^2-2 \alpha  \partial _zA_t \psi ^2 z+\partial _z\partial _xA_x+\partial
   _z\partial _yA_y-\partial _z\partial _zA_t=0,&\label{A2}\\
\frac{\partial _zA_x f'}{2}+\alpha  f \partial _zA_x \partial _z\psi  \psi  z^2+\alpha  f \partial _zA_x \psi ^2
   z+\frac{f \partial _z\partial _zA_x}{2}-\alpha  \partial _tA_x \partial _z\psi  \psi  z^2+\alpha  (-\partial
   _tA_x) \psi ^2 z-\partial _t\partial _zA_x-\alpha  \partial _t\psi  \partial _zA_x \psi  z^2\nonumber\\
   +\alpha  \partial
   _xA_t \partial _z\psi  \psi  z^2+\alpha  \partial _xA_t \psi ^2 z-\alpha  \partial _xA_y \partial _y\psi  \psi 
   z^2-\frac{\partial _x\partial _yA_y}{2}+\alpha  \partial _yA_x \partial _y\psi  \psi  z^2+\frac{\partial
   _y\partial _yA_x}{2}+\frac{\partial _z\partial _xA_t}{2}=0,&\label{A3}\\
\frac{\partial _zA_y f'}{2}+\alpha  f \partial _zA_y \partial _z\psi  \psi  z^2+\alpha  f \partial _zA_y \psi ^2
   z+\frac{f \partial _z\partial _zA_y}{2}-\alpha  \partial _tA_y \partial _z\psi  \psi  z^2+\alpha  (-\partial
   _tA_y) \psi ^2 z-\partial _t\partial _zA_y-\alpha  \partial _t\psi  \partial _zA_y \psi  z^2\nonumber\\+\alpha  \partial
   _xA_y \partial _x\psi  \psi  z^2+\frac{\partial _x\partial _xA_y}{2}-\frac{\partial _x\partial _yA_x}{2}-\alpha 
   \partial _x\psi  \partial _yA_x \psi  z^2+\alpha  \partial _yA_t \partial _z\psi  \psi  z^2+\alpha  \partial _yA_t
   \psi ^2 z+\frac{\partial _z\partial _yA_t}{2}=0,&\label{A4}\\
2 \alpha  f \partial _x\psi  \partial _zA_x \psi  z^2+2 \alpha  f \partial _y\psi  \partial _zA_y \psi  z^2+f \partial _z\partial _xA_x+f \partial _z\partial _yA_y-2 \alpha  \partial _tA_x \partial _x\psi  \psi  z^2-2 \alpha  \partial _tA_y \partial
   _y\psi  \psi  z^2-\partial _t\partial _xA_x\nonumber\\
-\partial _t\partial _yA_y-\partial _t\partial _zA_t-2 \alpha  \partial _t\psi  \partial _zA_t \psi  z^2+2 \alpha  \partial _xA_t \partial _x\psi  \psi  z^2+\partial _x\partial _xA_t
   +2 \alpha  \partial _yA_t
   \partial _y\psi  \psi  z^2+\partial _y\partial _yA_t=0.&\label{A5}
\end{align}

Eq.~(\ref{A5}) is a constraint equation, which in the conformal boundary is
\begin{align}
\partial _t\rho=-\partial _x\partial _xA_t-\partial _y\partial _yA_t-\partial _z\partial _xA_x-\partial _z\partial _yA_y.
\end{align}

For the system with $U(1)$ symmetry, the formulas are as follows
\begin{align}
-2 i A_t \partial _z\psi +A_x^2 \psi +2 i A_x \partial _x\psi +A_y ^2 \psi +2 i A_y \partial _y\psi +2 \partial _t\partial _z\psi +i \partial
   _xA_x \psi -\partial _x\partial _x\psi +i \partial _yA_y \psi -\partial _y\partial _y\psi-i \partial _zA_t \psi \nonumber\\
   +\partial _z\partial _z\psi 
   z^3-\partial _z\partial _z\psi +3 \partial _z\psi  z^2+3 \psi ^{*^2} \tau  \psi ^3 z^2+2 \lambda  \psi ^* \psi ^2+\psi  z=0,&\label{A1C}\\
   \partial _z\partial _xA_x+\partial _z\partial _yA_y-\partial _z\partial _zA_t-i \partial _z\psi ^* \psi +i \partial _z\psi  \psi ^*=0,&\label{A2C}\\
   2 A_x \psi ^* \psi +2 \partial _t\partial _zA_x+\partial _x\partial _yA_y-i \partial _x\psi ^* \psi +i \partial _x\psi  \psi ^*-\partial
   _y\partial _yA_x+3 \partial _zA_x z^2-\partial _z\partial _xA_t+\partial _z\partial _zA_x z^3-\partial _z\partial _zA_x=0,&\label{A3C}\\
   2 A_y \psi ^* \psi +2 \partial _t\partial _zA_y+\partial _x\partial _yA_x-i \partial _y\psi ^* \psi +i \partial _y\psi 
   \psi ^*-\partial _x\partial _xA_y+3 \partial _zA_y z^2-\partial _z\partial _yA_t+\partial _z\partial _zA_y z^3-\partial _z\partial _zA_y=0,&\label{A4C}\\
   -2 A_t \psi ^* \psi -\partial _t\partial _xA_x-\partial _t\partial _yA_y-\partial _t\partial _zA_t+i \partial _t\psi ^* \psi -i \partial _t\psi 
   \psi ^*+\partial _x\partial _xA_t+\partial _y\partial _yA_t-\partial _z\partial _xA_x z^3+\partial _z\partial _xA_x\nonumber\\
   -\partial _z\partial _yA_y
   z^3+\partial _z\partial _yA_y-i \partial _z\psi ^* \psi +i \partial _z\psi ^* \psi  z^3-i \partial _z\psi  \psi ^* z^3+i \partial _z\psi  \psi
   ^*=0.&\label{A5C}
\end{align}
Eq.~(\ref{A5C}) is a constraint equation, which in the conformal boundary is
\begin{align}
\partial _t\rho=-\partial _x\partial _xA_t-\partial _y\partial _yA_t-\partial _z\partial _xA_x-\partial _z\partial _yA_y.
\end{align}
Numerically, we employ the Chebyshev pseudospectral method in the holographic direction with $n_z = 21$ grid points. On the two-dimensional spatial plane, we use the Fourier spectral method, which imposes periodic boundary conditions on the system, with $n_x = n_y = 200$ grid points. For the time direction, we adopt the fourth-order Runge-Kutta method with a time step $\delta t = 0.05$.
%数值计算上，

\section{Controlled defect configuration}
\label{app:B}
% 重新设置公式编号为 A-1, A-2, ...
\renewcommand{\theequation}{B-\arabic{equation}}
\setcounter{equation}{0}
% 重新设置图形编号为 Figure A-1, Figure A-2, ...
\renewcommand{\thefigure}{B-\arabic{figure}}
\setcounter{figure}{0}

\begin{figure*}
    \centering
    \includegraphics[width=\columnwidth]{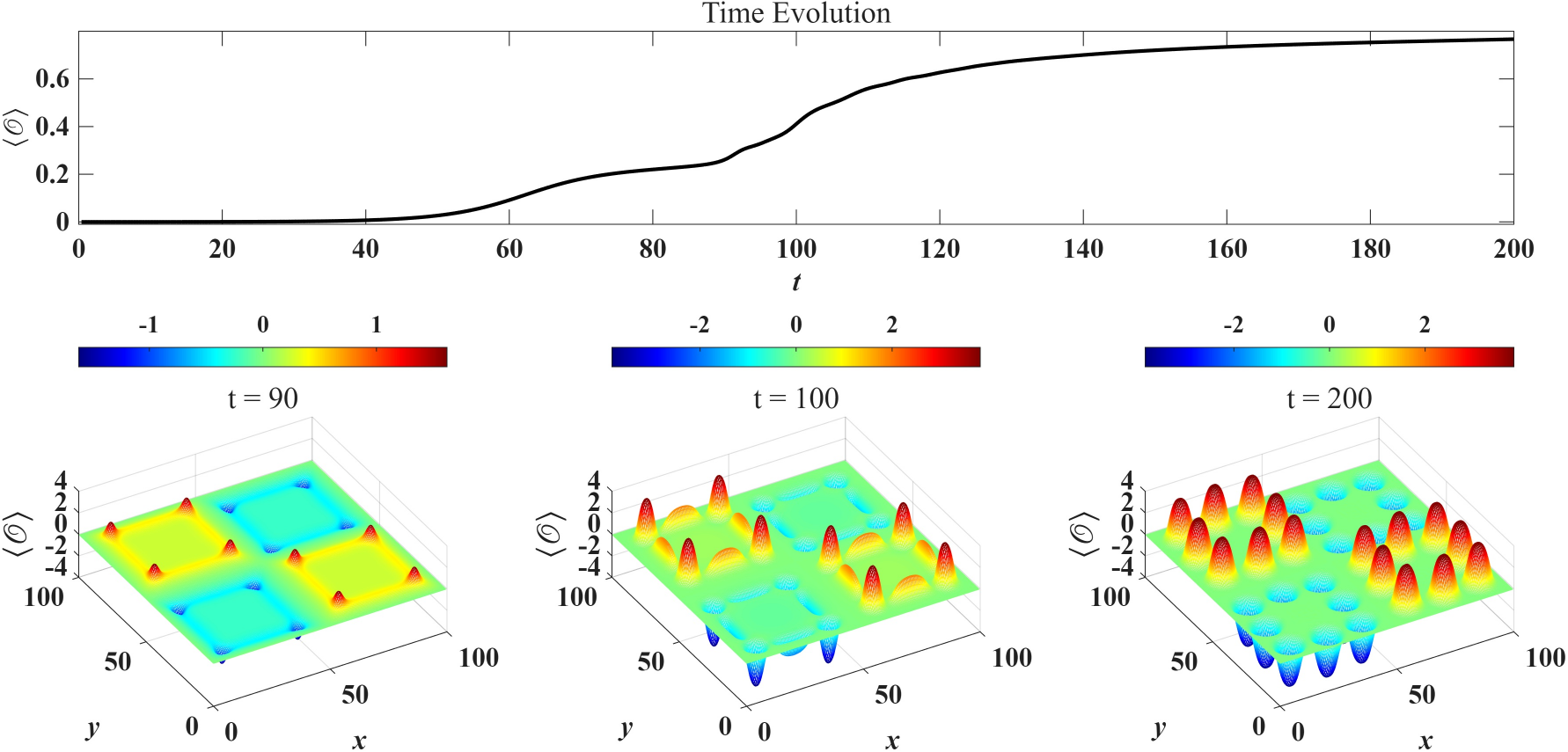}
    \caption{The time evolution of the scalar field on a two-dimensional plane with fixed initial configuration. The initial perturbation is $\psi_i(x,y) = \{-10^{-5},(0\leq x<L_x/2$ and $0\leq y<L_y/2)$ or $(L_x/2\leq x\leq L_x$ and $L_x/2\leq y\leq L_y);10^{-5},(L_x/2\leq x\leq L_x$ and $0\leq y\leq  L_y/2)$ or $(0\leq x<L_x/2$ and $L_y/2\leq y\leq L_y$. The upper panel shows the average condensate as a function of time. The lower panels display the spatial distribution of the expectation value at different times. The color bar indicates the magnitude of the expectation value. In this process, we performed only a single quench with $T_i = 1.041T_c$, $T_f = 0.963T_c$, $\lambda = -4$, and $\tau = 2.7$.}\label{PhaseS2DfixInitial}
\end{figure*}

To eliminate stochastic effects from random initial conditions, we further consider a controlled initial configuration with predefined domain-wall structures. The initial perturbation is divided into four regions with opposite signs, which deterministically generates domain walls near the interfaces. The subsequent evolution is shown in Fig.~\ref{PhaseS2DfixInitial}.

One can clearly observe that phase separation first develops near the intersections of the domain walls, where the local inhomogeneity is maximal. The nucleated regions then expand outward and invade the surrounding condensate. These results further confirm that topological defects act as deterministic seeds for phase separation. It is worth noting that in this process we do not adopt the double-quench method. That is, symmetry breaking and phase separation are not strictly separated into two distinct stages. Although phase separation always lags behind symmetry breaking, the two processes overlap to some extent in this single-quench scenario. This implies that their coupling effect may be more complex and may not be simply described by the cooperative effect discussed in this paper. Nevertheless, the evolution behavior in this case still exhibits a certain degree of induced effect.

\end{document}